\documentstyle[twoside, epsf]{article}

\input ibvs2.sty

\begin{document}
\IBVShead{5xxx}{00 Month 200x}

\IBVStitle{Outburst of a WZ~Sge-type dwarf nova, AL Com in 2007}

\IBVSauth{Uemura, M.$^1$; Arai, A.$^2$, Sasada, M.$^2$, Schmeer, P.$^3$, Miller, I.$^4$, Ohsugi, T.$^{1,2}$, Yamashita, T.$^1$, Kawabata, K. S.$^1$, Isogai, M.$^1$, Sato, S.$^5$, Kino, M.$^5$}

\IBVSinst{Astrophysical Science Center, Hiroshima University, Kagamiyama
1-3-1, Higashi-Hiroshima 739-8526, Japan. e-mail:
uemuram@hiroshima-u.ac.jp}
\IBVSinst{Department of Physical Science, Hiroshima University,
  Kagamiyama 1-3-1, Higashi-Hiroshima 739-8526, Japan}
\IBVSinst{Bischmisheim, Am Probstbaum 10, 66132 Saarbr\"{u}cken, Germany}
\IBVSinst{Furzehill House, Ilston, Swansea, South Wales, SA2 7LE, UK}
\IBVSinst{Department of Physics, Nagoya University, Furo-cho,
Chikusa-ku, Nagoya 464-8602, Japan}

\SIMBADobjAlias{AL Com}
\IBVStyp{UGWZ}
\IBVSkey{photometry}
\IBVSabs{We report photometric observations of AL Com during its rare outburst 
in 2007.  The light curve is reminiscent of its past superoutbursts in 1995 
and 2001, except for the rebrightening phase after the main superoutburst. 
During the rebrightening phase in 2007, we found clear modulations 
between $V=16.2$---$15.2$.  In conjunction with the lack of 
prominent superhumps in our time-series observations, the modulations can 
most naturally be interpreted as repetitive short rebrightenings with
a cycle of 1---2~days.  The rebrightening characteristics in 2007 are 
different from those in 1995 and 2001.  This indicates that the type
of rebrightenings in WZ~Sge stars depends not on binary parameters of
objects, but on the mass-accretion process for each outburst.}

\begintext

AL~Com is a WZ~Sge-type dwarf nova, which is a subclass of dwarf novae
characterized by very long recurrence times of outbursts.  In the case
of AL~Com, outbursts were observed in 1892, 1941, 1961, 1965, 1974,
1975 (Bertola 1964; Richter 1992), 1976, 1995 (Howell, et al., 1996;
Kato, et al., 1996; Patterson et al., 1996; Nogami, et al., 1997), and
2001 (Ishioka, et al., 2002).  Superhumps were detected for the first
time in 1995, and again in 2001.  The 1961, 1965, and 1975 outbursts 
also lasted more than 30~days, which were probably superoutbursts
(Richter 1992).  The light curve of its superoutbursts is
characterized by a ``dip'' which suddenly interrupts a plateau phase
of the superoutbursts (Bertola 1964; Richter 1992; Howell, et al.,
1996; Ishioka, et al., 2002). After the dip, AL~Com experienced a
rebrightening.  Several types of rebrightening phenomena have been
observed in WZ~Sge stars just after main superoutbursts (Richter 1992;
Kato, et al., 2004; Uemura, et al. 2007).  The mechanism of them and
the origin of their diversity are poorly understood.  The
rebrightening light curve of AL~Com is characterized by a long plateau
lasting more than 10~days. 

Here, we report a new outburst of AL~Com in October---November 2007.
We performed optical and near-infrared photometry at 3 observatories.
Details of our observational equipment are shown in table~1.  Using
the standard procedure of image reduction and aperture photometry, we
obtained magnitudes of AL~Com and comparison stars from our 
images.  As the optical comparison stars for the images obtained at
Higashi-Hiroshima and Ilston, we used a neighbor star located at  
12\hr 32\mm 10\fsec 04, $+$14\deg 20\arcm 15\farcs 3 with the AAVSO
$V$-band sequence ($V=13.509$, a star labeled as ``AUID~000-BBS-916'' 
\footnote{http://www.aavso.org/}).  For the infrared data obtained at
Higashi-Hiroshima, we used the same comparison star whose $J$-band
magnitude is quoted from the 2MASS catalog ($J=12.032$,
2MASS~12321003$+$1420153
\footnote{http://www.ipac.caltech.edu/2mass/}). 
For the optical data obtained at Iowa, we used a comparison star
located at 
12\hr 32\mm 05\fsec 30	$+$14\deg 23\arcm 34\farcs 0 with the $R_{\rm
  c}$-band sequence presented in Skiff~(2007) ($R_{\rm c}=13.09$,
labeled as ``NGC 4501 11'').  

Figure~1 shows the optical light curve of the outburst.  While our
observations are rather sparse due to a bad seasonal condition, the
feature of the light curve is reminiscent of the past superoutbursts
in 1995 and 2001; a main superoutburst until JD~2454405 and a
subsequent rebrightening phase until about JD~2454425.  We, hence, 
propose that this outburst is a superoutburst.  On the basis of the
latest 3 superoutbursts, the supercycle of AL~Com is calculated to
be $\sim 6\,{\rm yr}$.  This is the shortest among WZ~Sge stars
(Kato, et al., 2004), while the stability of the cycle should be
checked by a long monitoring in future.

A noteworthy feature of the 2007 superoutburst is the behavior during the 
rebrightening phase.  As can be seen in figure~1, the magnitude apparently 
oscillates in a range of $V=16.2$---$15.2$ 
between JD~2454410 and 2454421.  A clear short flare was, furthermore,
observed on JD~2454425, just before the final fading stage.  
These large amplitude variations were not seen during the past
rebrightenings of AL~Com, in which the object exhibited 
only low amplitude ($\sim 0.1\,{\rm mag}$) superhumps (Nogami, et al.,
1997).  The lower panel of figure~1 presents the color variation of
$V-J$.  The color became bluer when the object was brighter.  This is
a typical behavior of dwarf nova outbursts, suggesting an appearance
and disappearance of a hot, optically-thick accretion disk.  We note
that the color $V-J$ is atypically red during the rebrightening phase,
compared with typical colors at the maximum of dwarf nova outbursts
($V-J\sim 0$).  

In order to find possible superhumps, we performed time-series observations 
during the rebrightening phase.  The light curves are shown in
figure~2.  The figure contains 4 sets of light curves, in each of which the
left panel includes all observations and the right panel is a 
phase-averaged light curve using the superhump period of 0.05722~d
(Kato, et al., 1995).  As can be seen in these figures, we cannot find
significant periodic variation having amplitudes larger than 
$\sim 0.1\,{\rm mag}$.  The observed large oscillation is, hence, not
attributed to superhumps.  In conjunction with the color behavior, 
we conclude that the apparent oscillation is a sign of repetitive
short rebrightenings with a cycle of 1---2~days, as observed in
WZ~Sge (Patterson, et al.,~2002).

As mentioned above, it is unclear what determines the rebrightening types 
in WZ~Sge stars.  In this paper, we revealed that AL~Com exhibits 
not only long plateau type rebrightenings, but also short repetitive ones.  
This is the second case that different rebrightening behaviors were
unambiguously observed in a WZ~Sge star; WZ~Sge itself exhibited no
major rebrightening in the 1946 superoutburst, while short repetitive 
rebrightenings were observed in the 1978 and 2001 superoutbursts
(Patterson, et al., 1981).  EG~Cnc also experienced a hint of different
types of rebrightenings (Kato, et al., 2004).  These facts indicate
that the type of rebrightenings depends not directly on physical
parameters of binaries and their components, for example, mass ratios
or the strength of magnetic fields, but on the mass-accretion process
for each outburst. 

\IBVSfig{10cm}{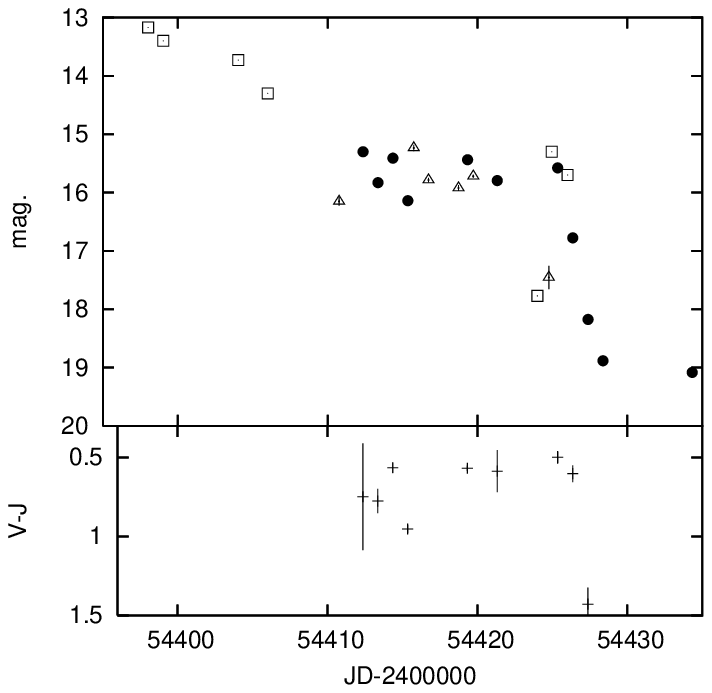}{Upper panel: Light curve of the 2007
 superoutburst of AL~Com.  The abscissa and ordinate denote the time
 in JD and the magnitude, respectively.  The filled circles are
 $V$-magnitudes obtained at Higashi-Hiroshima.  The open triangles and
 squares indicate unfiltered CCD observation at Ilston and Iowa.
 The magnitudes of Ilston's data were calculated by added the
 $V$-magnitude of the comparison star ($V=13.509$) to its differential 
 magnitude.  Those of Iowa's data were calculated by added the $R_{\rm
 c}$-magnitude of the comparison ($R_{\rm c}=13.09$) to its
 differential magnitude.  Errors of magnitudes are indicated as
 vertical bars, while most of errors are smaller than the symbol size. 
 Lower panel: Color variations.  The ordinate denotes $V-J$.}

\IBVSfig{4cm}{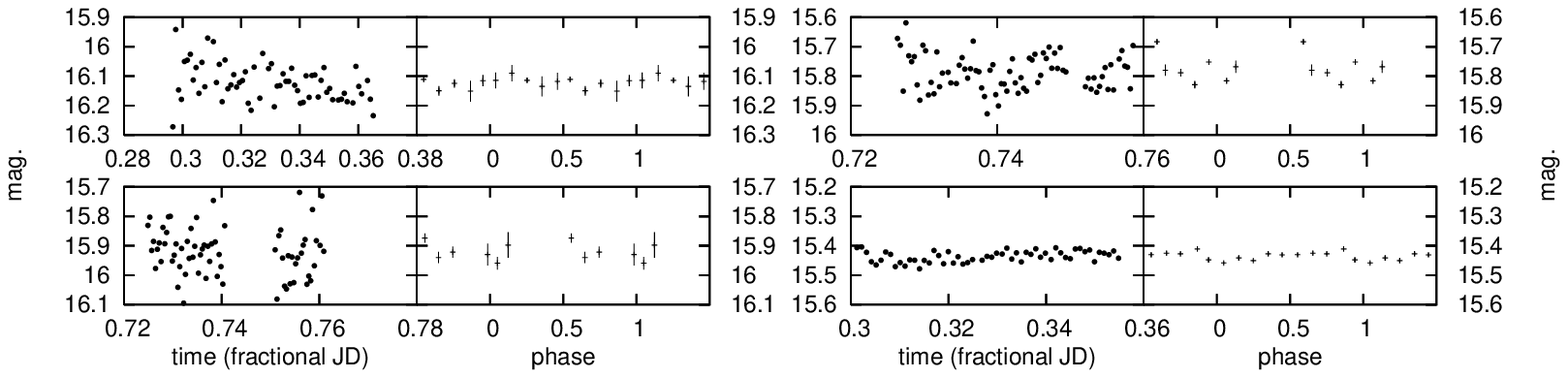}{Time-series light curves during the rebrightening 
 phase.  Observations were performed on JD~2454415 (upper left),
 2454416 (upper right),  2454418 (lower left), and 2454419 (lower
 right).  Each panel contains two light curves; the left ones show all
 data points and the right ones are phase-averaged light curves folded
 by the superhump period of 0.05722~d (Kato, et al., 1995).}

\centerline{Table 1. Details of instruments used for our observations.}
\vskip 3mm
\begin{center}
\begin{tabular}{ccccc}
\hline
Site & Telescope & Camera & Filter & Exposure time (sec) \\ 
\hline
Higashi-Hiroshima & 1.5-m (KANATA) & TRISPEC & $V$, $J$ & 63($V$), 60($J$) \\
Ilston & 35-cm & SXVF-H16 & no filter & 30 \\
Iowa & 37-cm (Rigel) & FLI SITe-003 & no filter & 25 \\
\hline
\end{tabular}
\end{center}

\vspace{2cm}

We appreciate useful comments and discussions about this paper by
T. Kato. This work was partly supported by a Grand-in-Aid from the
Ministry of Education, Culture, Sports, Science, and Technology of
Japan (17684004, 17340054, 18840032, 19740104).  

\references

Bertola, F., 1964, {\it Annales d'Astrophysique}, {\bf 27}, 298

Howell, S. B., Deyoung, J., Mattei, J. A., Foster, G., Szkody, P., 
 \& Cannizzo, J. K., 1996, {\it AJ}, {\bf 111}, 2367

Ishioka, R.,  Uemura, M.,  Matsumoto, K.,  Ohashi, H.,  Kato, T.,  Masi, G.,  
 Novak, R., et al., 2002, {\it A\&A}, {\bf 381}, L41

Kato, T., Nogami, D., Baba, H., Matsumoto, K., Arimoto, J., Tanabe, K., 
 Ishikawa, K., 1996, {\it PASJ}, {\bf 48}, L21

Kato, T., Nogami, D., Matsumoto, K., Baba, H., 2004, {\it PASJ}, {\bf 56}, S109

Nogami, D., Kato, T., Baba, H., Matsumoto, K., Arimoto, J., Tanabe, K., 
 Ishikawa, K., 1997, {\it ApJ}, {\bf 490}, 840

Patterson, J., McGraw, J. T., Coleman, L., \& Africano, J. L., 1981, 
 {\it ApJ}, {\bf 248}, 1067

Patterson, J., Augusteijn, T., Harvey, D. A., Skillman, D. R., Abbott,
	T. M. C., Thorstensen, J., 1996, {\it PASP}, {\bf 108}, 748

Patterson, J., Masi, G., Richmond, M. W., Martin, B., Beshore, E., 
 Skillman, D. R., Kemp, J., Vanmunster, T., 2002, {\it PASP}, {\bf 114}, 721

Richter, G. A., 1992, {\it Vina del Mar Workshop on Cataclysmic
  Variable Stars}, ed. by Nikolaus Vogt, ASP Conference Series (ASP:
  San Francisco),29, 12

Skiff, B. A., 2007, VizieR On-line Data Catalog: II/277.
	
Uemura, M., Arai, A., Krajci, T., Pavlenko, E., Shugarov, S. Yu., 
 Katysheva, N. A., Goranskij, V. P., Maehara, H., et al., 2008, {\it PASJ}, 
 accepted (arXiv:0711.2831)
 
\endreferences

\IBVSedata{5xxx-t2.txt}
\IBVSedata{5xxx-t3.txt}

\IBVSefigure{5xxx-f2.ps}
\IBVSefigure{5xxx-f3.ps}
\IBVSefigure{5xxx-f4.ps}
\IBVSefigure{5xxx-f5.ps}

\end{document}